\documentclass[a4paper,pra,preprint]{revtex4}
\usepackage{graphicx}
\usepackage{amsmath}
\usepackage{amssymb}
\usepackage{bm}

\begin{document}
\title{Cranked Hartree-Fock-Bogoliubov Calculation for 
Rotating Bose-Einstein Condensates}
\author{Nobukuni Hamamoto}
\altaffiliation{%
Present Address, Integrated Information Processing Center, Niigata University\\
8050 Ikarashi 2-no-cho,  Niigata, 950-2181, Japan\\
}
\email{hamamoto@cc.niigata-u.ac.jp}
\author{Makito Oi}
\affiliation{%
Department of Physics, University of Surrey\\
Guildford, Surrey, GU2 7XH, United Kingdom
}
\author{Naoki Onishi}
\affiliation{%
Department of Information System, Tokyo International University\\
1-13-1 Matoba-kita, Kawagoe, Saitama, 350-1197, Japan
}
\date{\today}
 
\begin{abstract}
A rotating bosonic many-body system in a harmonic trap is studied with
the 3D-Cranked Hartree-Fock-Bogoliubov method at zero temperature, 
which has been applied to nuclear many-body systems at high spin.
This method is a variational method extended from the Hartree-Fock theory,
which can treat the pairing correlations in a self-consistent manner.
An advantage of this method is that a finite-range interaction
between constituent particles can be used in the calculation,
unlike the original Gross-Pitaevskii approach.
To demonstrate the validity of our method, 
we present a calculation for a toy model,
that is, a rotating system of ten bosonic particles
interacting through the 
repulsive quadrupole-quadrupole 
interaction in a harmonic trap.
It is found that the yrast states, the lowest-energy states for
the given total angular momentum, does not correspond to 
the Bose-Einstein condensate, except a few special cases.
One of such cases is a vortex state,  which appears
when the total angular momentum $L$ is twice the particle number $N$
(i.e., $L=2N$).

\end{abstract}

\maketitle

\section{Introduction}

Since the Bose-Einstein condensation (BEC) was realized in trapped 
dilute atomic gases at ultra-low temperature, 
theoretical studies of the BEC have been rapidly developing. 

In the early stage of study, 
ultra-cold alkali atoms such as $^{87}$Rb and $^{23}$Na
were mainly used for a formation of the BEC. 
Many theoretical analyses were performed with 
the Gross-Pitaevskii (GP) equation, 
in which the two-body interaction takes a delta-function form.
In fact, the two-dimensional GP equation is suitable particularly
for the study of the BEC made of alkali atoms in a cylindrical trap,
where the s-wave scattering is dominant.

Thanks to recent experimental developments, 
non-alkali atoms and molecules were also cooled down to form the BEC.
Such an example is seen in the condensate of Cr atoms \cite{Crexpt}.
The inter-atomic potential in the BEC made of Cr atoms cannot
be approximated exclusively by the delta function because of 
the strong dipole moment carried by a Cr atom \cite{CRS05,ZZ05}.
Therefore, the dipole-dipole interaction (including a tensor force)
could be also responsible for the many-body dynamics, which involves
the d-wave scattering in the BEC.
Another example is the condensate of molecules \cite{BEC-BCS}, where
the delta function is not appropriate for the intermolecular interaction 
because of the anisotropic nature of the interaction.

The BEC can be rotated
with the state-of-the-art experimental techniques, 
such as the laser spoon \cite{Ke}.
Owing to these techniques, 
it was demonstrated that the BEC undergoes the quantum phase transition
to vortex states.
The ``cranked'' GP equation was applied to the analysis of the rotating BEC,
and it successfully explained the quantum phase transition, including
the triangular lattice of vortices \cite{KMP00,Triangle}.
In the early stage of the study of the rotating BEC, 
the two-dimensional GP equation was mainly
used for theoretical analyses \cite{GP2}, 
because the BEC was only rotated about the fixed axis. 

Recently observed phenomena, such as
precession \cite{expt-tilt-1,expt-tilt-2} and bending \cite{bendexpt} 
of vortices, require the rotational axis to move around 
in a time-dependent manner, with respect to a certain coordinate frame.
These phenomena attract much interest in terms of the three-dimensional 
spatial structures of the vortex states.
Hence, the three-dimensional GP equation is now being applied to 
various vortex states \cite{GP3-1,GP3-2,GP3-3,GP3-4,GP3-5}.
In addition, a topological technique 
was developed to produce vortices with spin 2 or 4 ($\hbar$),
by reversing the magnetic field of the trap \cite{BVortexRb,BVortexNa}. 
(Below, we take the unit for angular momentum to be $\hbar=1$.)
The formation mechanism of such vortices 
requires three-dimensional motion of the vortices.
In these situations, the total angular momentum vector needs to be treated
in a three-dimensional manner.

In this way, everytime new experimental progresses are achieved, 
new physical situations are created, to
which the original GP equation cannot be applied in a naive way.
The cranked Hartree-Fock-Bogoliubov (cranked HFB) method,
which has been used to describe rotational states of atomic nuclei 
\cite{3dCHFB,3dCHFB2},  could be a useful and powerful approach 
in order to deal with these new situations.

Rotating BEC systems in a trap has been analyzed
with the GP equation, or the exact diagonalization method 
using a set of the truncated basis.
Mottelson was the first to discuss the ``yrast'' structure \cite{BM75}. Here, 
the yrast states mean the lowest-energy states for given angular momentum.
He proposed a scenario \cite{Mottelson-1} 
that the quadrupole excitation is dominant when 
the total angular momentum $L$ is much less than 
the particle number $N$ ($L<<N$), and that
all the bosonic particles will occupy the p state 
when the total angular momentum becomes equal to the particle number.
This situation can be interpreted as a creation of a vortex state.
This prediction was numerically verified by himself and his collaborators
 using the two-dimensional GP equation \cite{Mottelson-2}. 
Bertsch and Papenbrock also verified this prediction
using the diagonalization of the two-dimensional model Hamiltonian
\cite{Bertsch-1,Bertsch-2}. 
Further detailed analysis was  performed by Nakajima et. al.\cite{Nakajima}.

In this paper, 
we apply the cranked HFB method to a simple schematic model, 
where bosonic particles interact weakly through the 
repulsive quadrupole-quadrupole interaction. 
%
This model is too simple to describe the detailed structure of   
realistic systems.
But, in limited situations, the quadrupole-quadrupole interaction  
becomes a phenomenologically valid interaction that can reflect physically  
essential properties of the realistic interaction.
For example, our model can describe low-energy rotational  
excitations of weakly interacting alkali atoms, as discussed by
Mottelson \cite{Mottelson-1}.
%
Then, using the density matrix of the yrast states, as well as
its eigenvalues and eigenstates, we compare our results with the other methods.
Creation of a vortex state is also discussed within the framework of our model.

In Sections II and III, we  present how the cranked HFB theory is extended
so as to calculate not only fermionic systems but also bosonic ones. 
Unlike the GP equation, we do not assume the inter-atomic
potential to be the delta function. Also, we do not
suppose an {\it a priori} existence of the condensate, 
which is the essential assumption in the GP equation.  
The cranked HFB theory is a constrained mean-field theory, 
and the value and direction of the total angular momentum vector
are controlled in the calculation.
With this method,
it is expected that we can numerically analyze 
not only structure of dilute many-boson systems in a trap, 
but also superfluidity produced by ultra-cold many-fermion systems in a trap.
In the present study, we focus on the study of weakly interacting Bose systems,
and an application of the cranked HFB theory to a simple model is presented
in Sections IV and V.

\section{Cranked Hartree-Fock-Bogoliubov Theory}
We describe the cranked HFB method that can be applied to rotating particles 
interacting two-body interactions.
This method was originally proposed for a description of nuclear rotation
\cite{3dCHFB,3dCHFB2}, but we extend the method to deal with 
not only fermions but bosons. 

Let $c_\alpha^\dagger, c_\alpha$ be the creation and annihilation
 operators of the single particle state 
$\langle\xi|\alpha \rangle =\psi_\alpha (\xi)$, where $\xi$
represents the real-space coordinates, 
spin coordinates and nuclear spin of particles.
The creation and annihilation operators of quasi-particle 
$a^\dagger_i, a_i$ is given by the
Bogoliubov-Valatin transformation \cite{Bogo,Valatin},
\begin{eqnarray}
a_i^\dagger &=& \sum_\alpha 
U_{\alpha i}   c_\alpha^\dagger +V_{\alpha i}   c_\alpha, \\
a_i         &=& \sum_\alpha 
U_{\alpha i}^* c_\alpha         +V_{\alpha i}^* c_\alpha^\dagger.
\label{BV-trans}
\end{eqnarray}
The operators $a_i,c_\alpha$ obey the following commutation rule,
\begin{eqnarray}
\label{comm-rule}
[ a_i, a_j^\dagger ]_{\pm} &=& \delta_{ij}, \
[ a_i^\dagger, a_j^\dagger ]_{\pm} =[ a_i, a_j ]_\pm = 0,\\ \nonumber
\left[c_{\alpha}, c_{\beta}^{\dag}\right]_{\pm} &=& \delta_{\alpha \beta}, \
[ c_{\alpha}^\dagger, c_{\beta}^\dagger ]_{\pm} =[ c_{\alpha}, c_{\beta} ]_\pm = 0,
\end{eqnarray}
where the upper sign ($+$) applies to fermions and the lower ($-$) to bosons.
To satisfy the commutation rule (\ref{comm-rule}), 
we need the following relations.
\begin{eqnarray}
U^\dagger U   \pm V^\dagger V   =1, U^T U^* \pm V^T V^* =1, \\ 
U^\dagger V^* \pm V^\dagger U^* =0, V^T U   \pm U^T V   =0.
\end{eqnarray}
Based on the variational principle, the $U$ and $V$ are determined.
The variational ansatz is chosen to be 
\begin{equation}
|\Phi\rangle = N_f \exp(\sum_{\alpha \beta}
\frac{1}{2} f_{\alpha \beta} c^\dagger_\alpha c^\dagger_\beta) |0 \rangle, 
\quad
f_{\alpha \beta} = \sum_i V_{\alpha i} ({U^*}^{-1})_{i\beta},
\label{HFBansatz}
\end{equation}
where, $N_f$ is a normalization constant and $|0\rangle$ is the true vacuum.
The variational state $|\Phi\rangle$ corresponds to the vacuum 
in the quasi-particle basis, that is, $a_i |\Phi \rangle =0$.
The many-body Hamiltonian including a two-body interaction
$V(\xi_1,\xi_2)$ is generally written as
\begin{equation}
\label{hamiltonian}
\hat{H} =
\sum_{\alpha \beta} H^0_{\alpha \beta}c^\dagger_\alpha c_\beta
+\frac{1}{4}
\sum_{\alpha \beta \gamma \delta}
{\cal V}_{\alpha \beta \gamma \delta}
 c^\dagger_\alpha
 c^\dagger_\beta
 c_\delta
 c_\gamma,
\end{equation}
where ${\cal V}_{\alpha\beta\gamma\delta}$ is given by
\begin{widetext}
\begin{equation}
{\cal V}_{\alpha \beta \delta \gamma}
   = \langle \psi_\alpha(\xi_1) \psi_\beta(\xi_2)   |V(\xi_1,\xi_2)|
             \psi_\gamma(\xi_1) \psi_\delta(\xi_2)) \rangle
 \mp \langle \psi_\alpha(\xi_1) \psi_\beta(\xi_2)   |V(\xi_1,\xi_2)|
             \psi_\delta(\xi_1) \psi_\gamma(\xi_2)) \rangle.
\end{equation}
\end{widetext}
The one-body part $H^0_{\alpha \beta}$ includes the kinetic energy and 
the spherical confinement potential.
Using Wick's theorem, we can represent 
the Hamiltonian (\ref{hamiltonian}) by 
the total energy $E$ and the one-body Hamiltonian $\hat{h}$, as
\begin{widetext}
\begin{eqnarray}
\hat{H} &=& E + \hat{h} +\frac{1}{4}
\sum_{\alpha \beta \gamma \delta}
{\cal V}_{\alpha \beta \gamma \delta}
: c^\dagger_\alpha
 c^\dagger_\beta
 c_\delta
 c_\gamma
:,\\
E&=&
\sum_{\alpha \beta} H_{\alpha \beta}^0 
\langle c_\alpha^\dagger c_\beta \rangle
+
\frac{1}{4}\sum_{\alpha \beta \gamma \delta}
{\cal V}_{\alpha \beta \gamma \delta}
( 
 \langle c_\delta c_\gamma \rangle \langle c^\dagger_\alpha c^\dagger_\beta \rangle
+\langle c^\dagger_\alpha c^\dagger_\beta \rangle \langle c_\delta
c_\gamma \rangle
+2 \langle c^\dagger_\alpha c_\gamma \rangle \langle c^\dagger_\beta
c_\delta \rangle
),\\
\label{onebody}
\hat{h} &=& 
\sum_{\alpha \beta} H_{\alpha \beta}^0 
: c_\alpha^\dagger c_\beta :
+
\frac{1}{4}\sum_{\alpha \beta \gamma \delta}
{\cal V}_{\alpha \beta \gamma \delta}
( 
 \langle c_\delta c_\gamma \rangle : c^\dagger_\alpha c^\dagger_\beta :
+\langle c^\dagger_\alpha c^\dagger_\beta \rangle :c_\delta c_\gamma:
+2 \langle c^\dagger_\alpha c_\gamma \rangle : c^\dagger_\beta c_\delta:
),
\end{eqnarray}
\end{widetext}
where $:\cdots:$ is the normal order product with respect to
$a,a^\dagger$ and an abbreviated expression is introduced for
 an expectation value, 
$\langle {\cal O} \rangle=\langle \Phi |{\cal O}| \Phi \rangle$.

The HFB wavefunction $|\Phi \rangle$ is determined through 
the variational principle with constraints $\hat{C}_n$,
\begin{equation}
\delta \langle \Phi | \hat{H} - \sum_n \mu_n \hat{C}_n | \Phi \rangle =0,
\label{cond-vari}
\end{equation}
where $\mu_n$ is a Lagrange multiplier.
In this method, 
three components of the total angular momentum 
and the particle number are constrained.
Further constraints are imposed on the following quadrupole operators,
$
\hat{B}_1 = \sqrt{\frac{15}{2\pi}} yz,
\hat{B}_2 = \sqrt{\frac{15}{2\pi}} zx,
\hat{B}_3 = \sqrt{\frac{15}{2\pi}} xy$,
in order to fix the intrinsic coordinate axes of the system
along the principal axes of the quadrupole moments.
Therefore, we have seven constraints in our calculations.
\begin{eqnarray}
\langle \Phi | \hat{J}_x | \Phi \rangle &=& J_x,
\langle \Phi | \hat{J}_y | \Phi \rangle  =  J_y,
\langle \Phi | \hat{J}_z | \Phi \rangle  =  J_z,\\
\langle \Phi | \hat{B}_1 | \Phi \rangle &=& 0,
\langle \Phi | \hat{B}_2 | \Phi \rangle  =  0,
\langle \Phi | \hat{B}_3 | \Phi \rangle  =  0,\\
\langle \Phi | \hat{N} | \Phi \rangle   &=& N,
\end{eqnarray}
In Eq.(\ref{cond-vari}), these constraints are represented 
with $\hat{C}_{n}$'s as
$
\hat{C}_1=\hat{J}_x,
\hat{C}_2=\hat{J}_y,
\hat{C}_3=\hat{J}_z,
\hat{C}_4=\hat{B}_1,
\hat{C}_5=\hat{B}_2,
\hat{C}_6=\hat{B}_3,
\hat{C}_7=\hat{N}.
$
In particular, the term $-\sum_{n=1}^{3}\mu_n\hat{C}_n$ has been
called the ``cranking term'' in nuclear high-spin physics, because
it simulates the effect of the Coriolis force in the rotating mean-field 
system.

\section{Method of Steepest Descent}

As mentioned earlier, we determine the HFB states,
following the variational principle.
By multiplying a unitary operator
to an arbitrary  initial HFB state $|\Phi\rangle$, 
another HFB state $|\Phi'\rangle$ is obtained.
This transformation is considered as a variational procedure
with respect to the matrices $U$ and $V$ of the Bogoliubov-Valatin
transformation, Eq.(\ref{BV-trans}). 
The transformation is iterated until a local minimum is found
to satisfy Eq.(\ref{cond-vari}).

Now, let us explain how the unitary operator is given within our framework.
First of all, 
from the extended Thouless theorem \cite{Thouless,Onishi}, 
the unitary transformation of the HFB state is expressed as
\begin{equation}
|\Phi' \rangle = \exp(\hat{d}) |\Phi \rangle,
\end{equation}
where
$\hat{d}$ is an anti-Hermitian operator 
$\hat{d}=-\hat{d}^\dagger$, which is generally expressed as,
\begin{equation}
  \hat{d} = \frac{1}{2}\sum_{ij}\left(
    d_{ij}a_i^\dagger a_j^\dagger -d_{ij}^* a_j a_i
  \right).
\end{equation}
A quasi-particle basis is then transformed in the following way.
\begin{equation}
\left(
\begin{array}{c}
a_i'^\dagger\\
a_i'
\end{array}
\right)
=
\left(
\begin{array}{c}
e^{-\hat{d}} a_i^\dagger e^{\hat{d}}\\
e^{-\hat{d}} a_i         e^{\hat{d}}
\end{array}
\right)
=\exp({\pm{\cal D}})^T 
\left(
\begin{array}{c}
a_i^\dagger\\
a_i
\end{array}
\right), 
\end{equation}
where
\begin{equation}
{\cal D} =
\left(
\begin{array}{cc}
0  & d\\
d^*& 0
\end{array}
\right),
\end{equation}
and $d$ is a matrix representation of $\hat{d}$.
We choose the anti-Hermite operator $\hat{d}$ as,
\begin{equation}
\hat{d}=[ \eta \hat{r} +\hat{s}]^{\rm a},
\end{equation}
where we define $[\hat{\cal O}]^{\rm a}=
\frac{1}{2}[\hat{\cal N},\hat{\cal O}]$ and
the quasi-particle number operator is given 
as $\hat{\cal N}=\sum_i a_i^\dagger a_i$.
An operator $\hat{s}$ and the single-particle Routhian $\hat{r}$ are
respectively defined as 
\begin{eqnarray} 
  \hat{s} &=& \sum_n \delta_n \hat{C}_n,\\
  \hat{r} &=& \hat{h}-\sum_n \mu_n \hat{C}_n.
\end{eqnarray}

Then, these parameters $\eta$, $\delta_n$, and $\mu_n$ are determined through
a minimization of  $\langle \Phi' | \hat{H} | \Phi' \rangle$ 
under the constraints $\langle \Phi | \hat{C}_i | \Phi \rangle =c_i$.
The parameters $\delta_n$ and $\mu_n$ are evaluated by expanding
$\langle \Phi' | \hat{C}_i | \Phi' \rangle$
up to the first order in $\delta_n$ and $\mu_n$. That is,
\begin{eqnarray}
\delta_k &=& \sum_i L_{ki}^{-1}(c_i -\langle \Phi |\hat{C}_i |\Phi \rangle),\\
\mu_k &=&  \sum_i L_{ki}^{-1}\langle \Phi | [\hat{C}_i,[\hat{h}]^a]|\Phi \rangle,
\end{eqnarray}
where $L_{ki}=
\langle \Phi | [ \hat{C}_i,[ \hat{C}_k ]^a ] | \Phi \rangle$.
Whereas, the parameter $\eta$ is determined 
from a minimization condition for $E=\langle \Phi' | \hat{H} | \Phi' \rangle$,
through expanding $E$ up to the second order in $\eta$.
As a consequence, we have
\begin{equation}
\eta=
-\frac{
 \langle \Phi | [\hat{H},[\hat{r}]^a] | \Phi \rangle
+\langle \Phi | [[\hat{H},[\hat{r}]^a],[\hat{s}]^a] | \Phi \rangle
}
{
 \langle \Phi | [[\hat{H},[\hat{r}]^a],[\hat{r}]^a] | \Phi \rangle
}.\end{equation}

To check the convergence for the self-consistency in the calculation, 
it is convenient to define the norm of $d$ 
as $|d|=\sqrt{\sum_{ij}\frac{1}{2}|d_{ij}|^2}$.
In our calculations, a criterion for the convergence is given whether
$|d|$ is less or greater than $\epsilon=1.0\times10^{-7}$.
When $|d| < \epsilon$, we judge that the convergence is numerically achieved.
Otherwise, the iteration for the self-consistency continue 
until $|d|$ meets the above condition.

\section{A schematic model}
To examine the convergence procedure of our method (3D-cranked HFB)
for the bosonic case,
let us consider a toy model, which is similar to
a realistic system of dilute ultra-cold Bose gases confined by 
an isotropic harmonic oscillator potential $V(r)=\frac{1}{2}M \omega^2 r^2$,
where $M$ represents the atomic mass.
We choose the single-particle states to be the harmonic oscillator states,
$\langle\bm{r}|\alpha \rangle = 
\langle\bm{r}|c^\dagger_{\alpha} |0\rangle = R_{n_\alpha l_\alpha}(r) 
i^{l_\alpha} Y_{l_\alpha m_\alpha}(\theta \phi)$, that is, a product of the
Laguerre polynomial and the spherical harmonics.
Let us denote this basis as $\alpha \equiv (n_{\alpha},l_{\alpha},m_{\alpha})$.

Mottelson discussed that the quadrupole correlation is important
in the low angular momentum region \cite{Mottelson-1}. 
In accordance with his proposition,
the following Hamiltonian is considered in our calculations.
\begin{eqnarray}
\hat{H}_{\rm model}&=& 
\hat{H}_0
+\frac{1}{2}\kappa \sum_{\mu=-2}^{2}(-)^\mu \hat{Q}_{-\mu} \hat{Q}_{\mu}
+g \hat{P}^\dagger \hat{P} \label{modelhamilt},\\
\hat{H}_0&=& 
\sum_\alpha (2n_\alpha +l_\alpha + \frac{3}{2})\hbar \omega \delta
c^\dagger_\alpha c_\alpha,\\
\hat{Q}_{\mu} &=& \sum_{\alpha \beta} 
\langle \alpha | 2 r^2 C_\mu^{(2)} | \beta \rangle
 c_\alpha^\dagger c_\beta, \\
 \hat{P} &=&
 \sum_\alpha \sqrt{2l_\alpha +1} 
 \langle l_\alpha m_\alpha  l_\alpha -m_\alpha | 00 \rangle
 c_{\bar{\alpha}} c_\alpha,
\end{eqnarray}
where $ c_{\bar{\alpha}}$ is the annihilation operator corresponding to
the state $\bar{\alpha}=(n_\alpha, l_\alpha, -m_\alpha)$,
and 
$C^{(k)}_{\kappa}(\Omega)$ is related to the spherical harmonics through
$C^{(k)}_\kappa(\Omega)=\sqrt{\frac{4\pi}{2k+1}}Y_{k \kappa}(\Omega)$.
This Hamiltonian should be considered 
as a simple model for weakly interacting dilute atomic gases.
%
%
The parameter $\kappa$ represents strength of the quadrupole-quadrupole  
interaction,
and $\kappa$ is positive in this work, to treat the repulsive two-body  
interaction.
The last term of Eq. (\ref{modelhamilt}) is called the pairing interaction.
\cite{Ring-Schuck} 
The parameter g represents the  
strength of the pairing
interaction. In the present calculation, we set g to be zero, for the  
sake of simplicity.
%
The one-body Hamiltonian (\ref{onebody}) in this model is given as,
\begin{equation}
\hat{h}_\text{model}= \hat{H}_0 +\kappa \sum_\mu (-)^\mu 
\langle \hat{Q}_{-\mu} \rangle \hat{Q}_\mu
+g \langle \hat{P}^\dagger \rangle \hat{P}.
\label{eq-scf}
\end{equation}

The oscillator energy for the isotropic harmonic oscillator states,
$\hbar \omega$, is set to 1.9 (meV). 
The quadrupole-quadrupole
interaction $\kappa/A^2$ is repulsive and its strength is set to 
0.1, 0.5 and 1.0 (meV/$\mu$m$^4$), where $A$ is the mass number of an atom.
%
%
The single-particle model space used in this calculation are
the harmonic oscillator states of the 
0s, 1s, 0d, 2s, 1d, 0g, 0p, 1p, 0f, 2p, 1f and 0h states.

To prepare the initial state, we use the deformed quadrupole mean field.
\begin{widetext}
\begin{equation}
\hat{h}_{\rm deform} = \hat{H}_0 
- \frac{2}{3} M\omega^2 r^2 
\left(
\beta\cos\gamma C^{(2)}_0(\Omega) 
+\frac{1}{\sqrt{2}}\beta\sin\gamma(C^{(2)}_2(\Omega)
+C^{(2)}_{-2}(\Omega)) 
\right).
\label{eq-deform}
\end{equation}
\end{widetext}

The quadrupole parameters $(\beta,\gamma)$ are useful measures
to think about the shape of the many-body system.
$\beta$ is a measure for elongation or stretching, 
while $\gamma$ for triaxiality or deviation from axial symmetry.
For example, a spherical shape has $\beta=0$ and 
nuclear superdeformation has typically $\beta \simeq 0.6$.
The triaxial parameter $\gamma$ gives axial shapes 
when $\gamma=0^{\circ}$ and $\gamma=60^{\circ}$.
The former shape corresponds to the so-called ``prolate''shape, 
which is similar to a kiwifruit,
while the latter to ``oblate'' shape, similar to a mandarin orange.
By definition, a triaxial shape is invariant with respect to an 
operation: $\gamma \rightarrow \gamma+120^{\circ}$.

We diagonalize the Hamiltonian (\ref{eq-deform}) for
$(\beta,\gamma)=(0.01,0)$ and obtain 
the single-particle states. The initial state is created 
as a Hartree-Fock state, that is, all the single-particle levels are
occupied below the Fermi level.

Since the initial state is symmetric under rotation about the $z$-axis,
collective rotation about the $z$-axis is suppressed.
In other words, we cannot crank the state around the symmetry axis.
To induce angular momentum, we initially set the constraints of 
the total angular momentum to $(J_x, J_y, J_z)=(0.05, 0, 0)$.
As a next step, we tilt the angular momentum vector to 
$(J_x, J_y, J_z)=(0, 0, 0.1)$.
With this procedure, we can increase $J_z$ up to 20,
with a step, $\Delta J_z = 0.05$.

\section{Results and discussions}

\begin{figure*}
\includegraphics[scale=1.0]{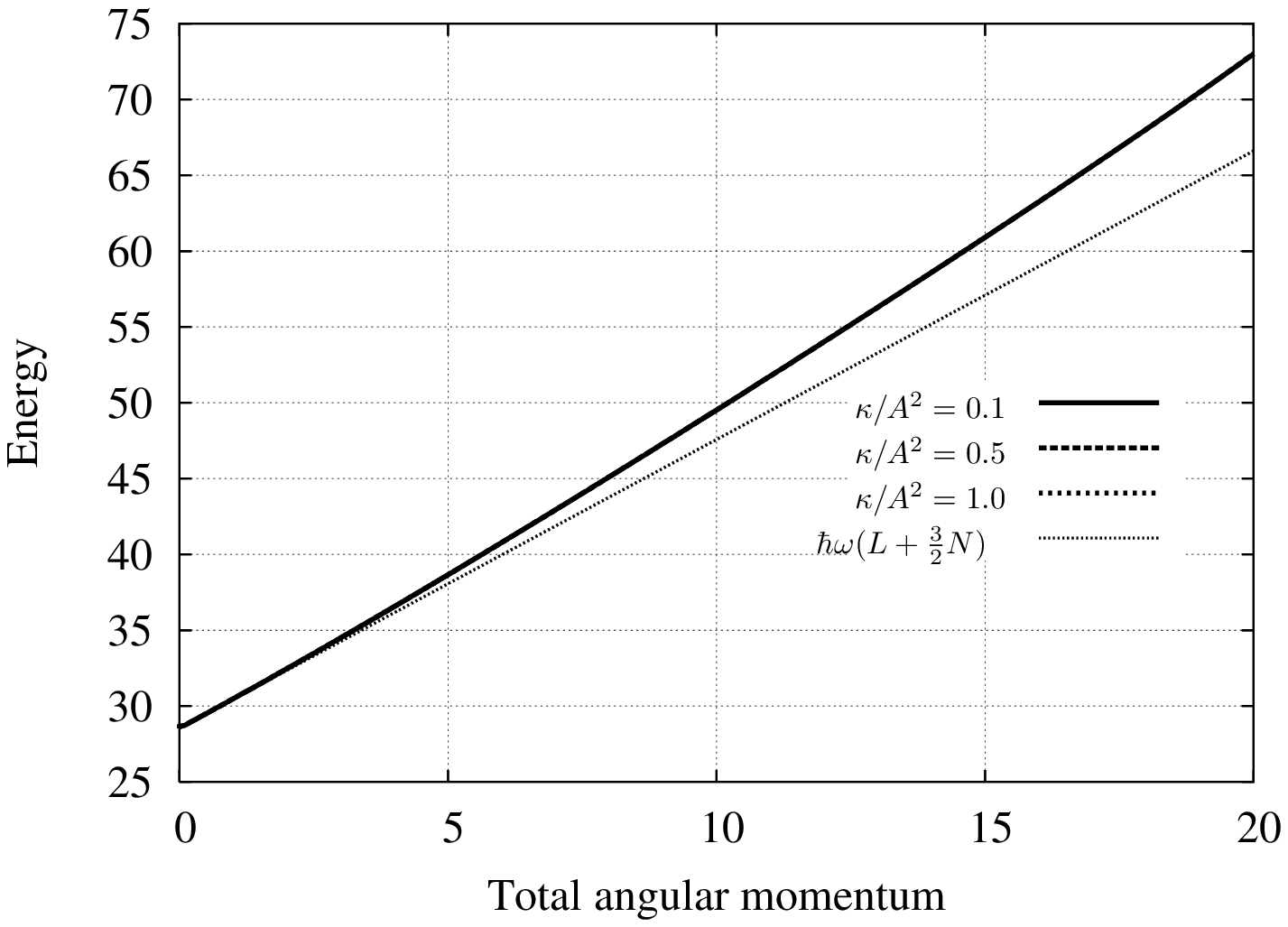}
\caption{The total energy (meV) as a function of the total angular momentum for
 $\kappa/A^2=0.1,0.5$ and $1.0$ (meV/$\mu$m$^4$).}
\label{fig:energy}
\end{figure*}

Figure \ref{fig:energy} represents the total energy $E$ 
as a function of the total angular momentum $L$. 
Four lines are plotted in this figure. 
One dotted line represents $\hbar\omega (L +\frac{3}{2}N)$
and the other three lines are calculated with the different
quadrupole-quadrupole interaction strengths, 
which are $\kappa/A^2=0.1, \ 0.5$ and $1.0$ (meV/$\mu$m$^4$).
Despite the different values for $\kappa$,
These lines are nearly identical.
In other words, $E$ is almost independent of $\kappa$.

We also find that $E$ increases almost in proportion to $L$, 
that is, $E\propto L$.
This result can be explained from a microscopic point of view.
As $L$ is increased, single-particle excitations are induced
one by one through the two-body interaction, so as to
satisfy the angular momentum constraints. In other words,
$\Delta E = \hbar\omega \Delta L$, where $\hbar\omega$ is the
single-particle energy spacing of the isotropic harmonic oscillator.
Due to the quantum statistics for bosons, 
this excitation mode can continue until the number of particles occupying
the ground state becomes zero.
This linear behavior is already noticed by other authors \cite{Bertsch-1}.
However, with a careful look at our numerical result, 
there is a slight deviation from the linearity
in the the total energy $(\hbar\omega(L+\frac{3}{2}N))$.
This small deviation is caused by deformation of the mean field,
which has a role to mix the single-particle orbits. 
The evolution of the quadrupole deformation in response to rotation
is discussed below.

\begin{figure*}
  \begin{tabular}{cc}
    \includegraphics[width=.45\textwidth]{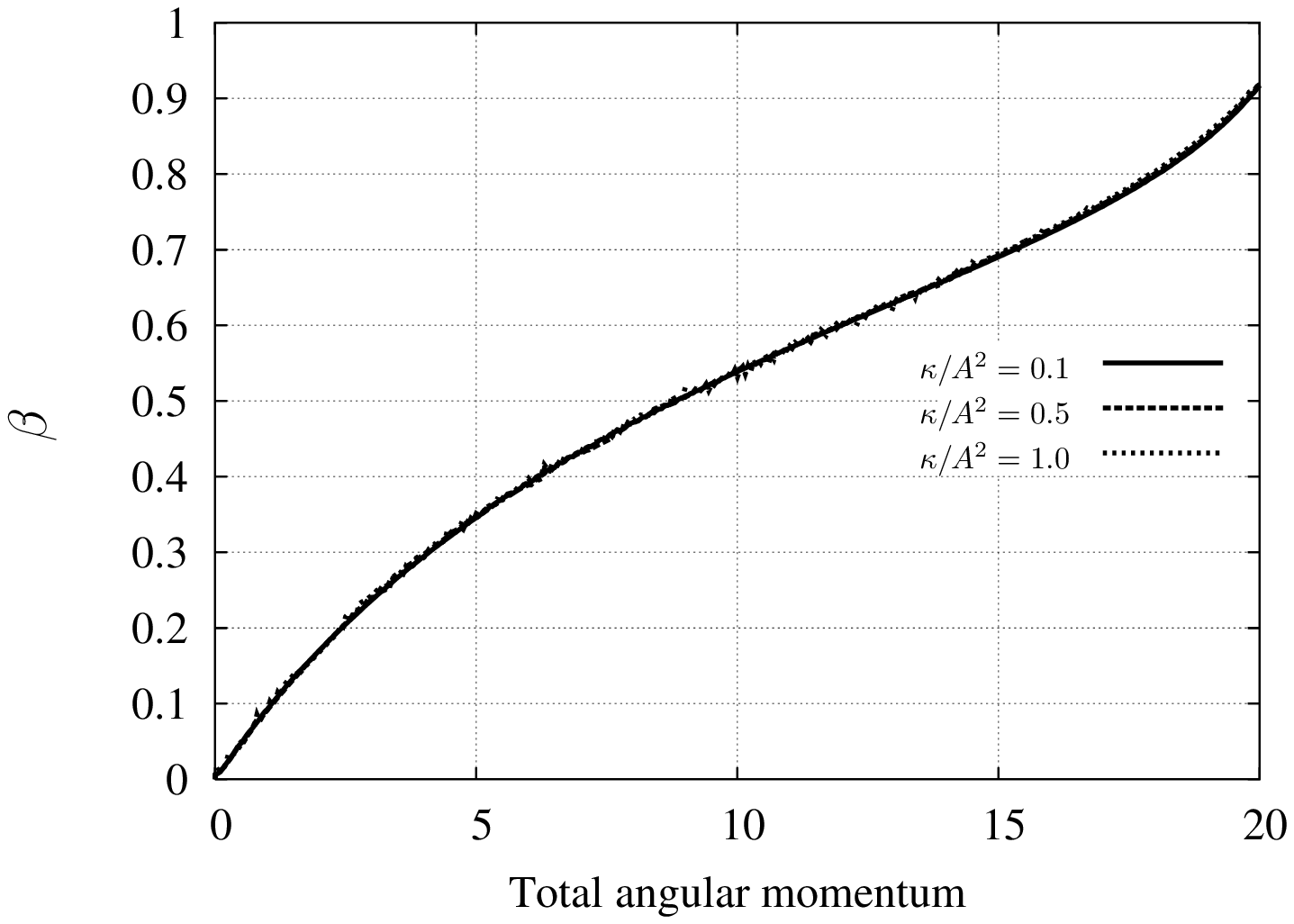}&
    \includegraphics[width=.45\textwidth]{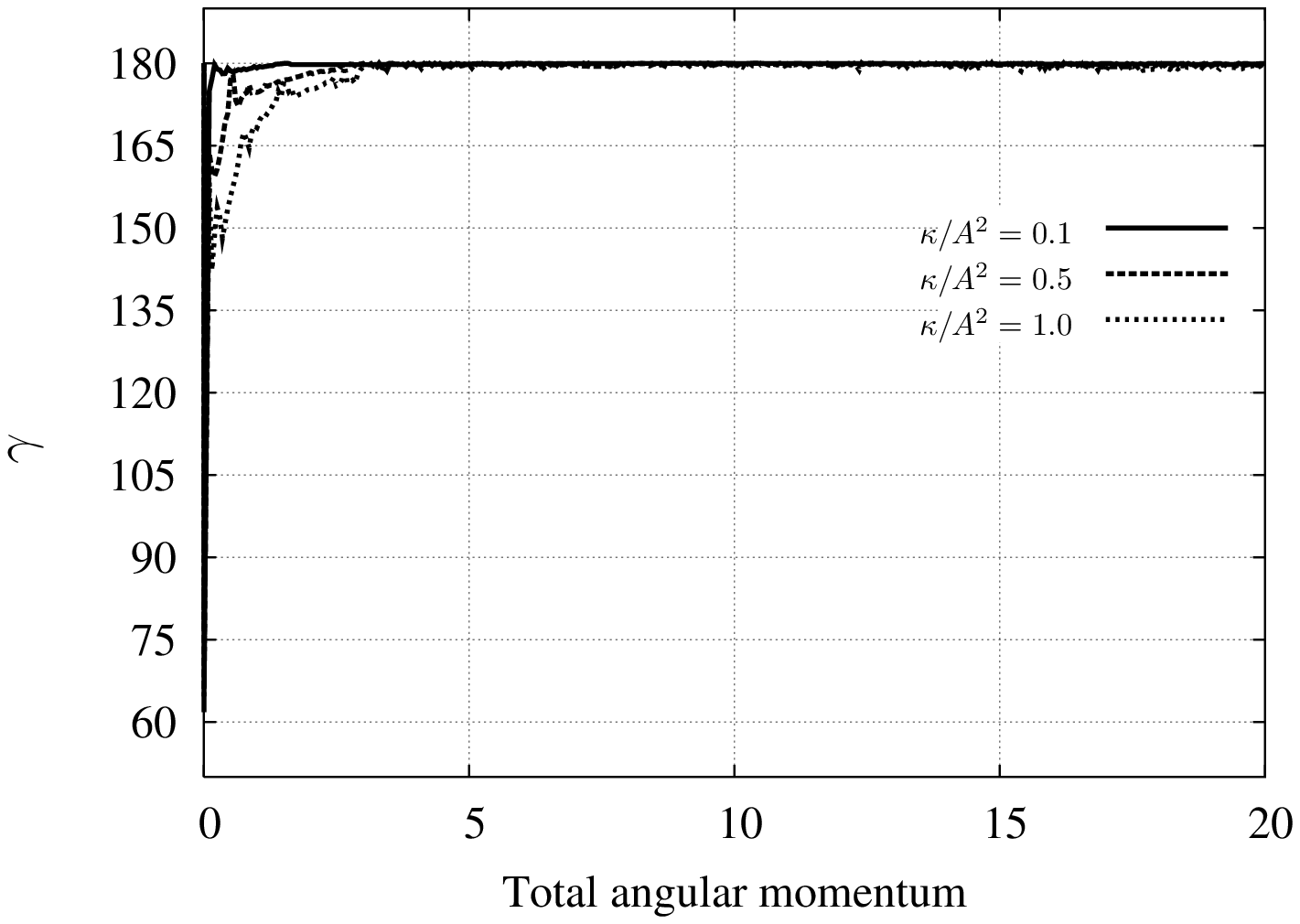}
  \end{tabular}
\caption{
The left and right panels represent the deformation 
parameters $\beta$ (unit-less) and $\gamma$ (degree) 
as a function of the total angular momentum, respectively.
}
\label{fig:betagamma}
\end{figure*}

Figure \ref{fig:betagamma} shows 
the deformation parameters $\beta$ and $\gamma$.
The strengths of the interaction are set in the same manner as in Figure 1.
These deformation parameters are self-consistently calculated as
\begin{eqnarray}
\beta &=& \frac{3\kappa}{M\omega^2}\sqrt{\langle Q_{20}\rangle^2 +2
 \langle Q_{22} \rangle^2},\\
\gamma &=& \tan^{-1} \frac{\langle Q_{22}\rangle}{\sqrt{2}\langle Q_{20} \rangle}.
\end{eqnarray}
The unit for the quadrupole moment ($M\omega^2/\kappa$) is derived
from the consistency at $L=0$ 
between the one-body Hamiltonian (\ref{eq-scf}) and
the deformed mean-field Hamiltonian (\ref{eq-deform}).
The figure shows that the deformation parameters
do not depend on the interaction strength very much,
although $\gamma$ shows minor differences at low spin.
When the total angular momentum is small ($L\alt 1$),
the mean field has a almost spherical shape.
This is because the trapping potential is spherical and
the present two-body interaction is repulsive.
As $L$ is increased, $\beta$  increases gradually.
In the small angular momentum region ($L < 5$ ),
$\gamma$ is not $180^{\circ} (\equiv 60^{\circ})$. 
(See the right panel of Figure \ref{fig:betagamma}.)
This result means that the shape of the mean field is not axial-symmetric,
but triaxial.
When the quadrupole-quadrupole interaction becomes stronger, 
the deformation tends to prefer a more triaxial-deformed shape.
However, any of the three cases ends up with the oblate shape 
($\gamma=180^{\circ}$) at high angular momentum ($L\agt 5$).
A reason for this tendency can be explained as the following:
When the quadrupole-quadrupole interaction is strong, 
the harmonic oscillator states having the different magnetic quantum numbers
become more mixed through the interaction.
As a result, the magnetic quantum number is no longer a good quantum number.
This is nothing but axial symmetry breaking, or an emergence of triaxiality.
It should be noted, however, that $\gamma$ is substantial only at low
angular momentum, where $\beta$ is very small. 
In other words, when the elongation is small ($\beta\simeq0$),
the triaxial degree of freedom is irrelevant 
in terms of a deviation from a spherical shape.
That is, in our calculation, 
the shape of the mean field can be regarded to be almost spherical 
in the small $L$ region.
On the other hand, for the higher angular momentum region ($L > 5$), 
$\gamma$ is almost constant to be $180^{\circ}$, meaning
that the mean field becomes an oblate shape.

The condition for the BEC of weakly interacting  bosonic atoms in a trap 
is given by $Nv/\hbar\omega << 1$ \cite{Mottelson-1}, 
where $v$ is an expectation value of the two-body interaction, 
while $\hbar\omega$ represents the single-particle level spacing.
In our calculation, $Nv$ corresponds to an expectation value
of the quadrupole-quadrupole force, that is, $\kappa 
(
\langle Q_{20} \rangle^2
+2\langle Q_{22} \rangle^2
)
$.
The ratio $Nv/\hbar\omega$
is then estimated to be $4.4\times 10^{-3} \beta A^2/\kappa$.
According to our calculation, deformation is up to $\beta\alt 1$, 
so that the ratio is of order of $10^{-4}$ to $10^{-5}$ for our
three choices of the interaction strength ($\kappa/A^2=0.1, \ 0.5, \ 1$
meV/$\mu$m$^4$).
This result means that our calculations can be regarded as
a weakly interacting many-boson system.

\begin{figure*}
\includegraphics[angle=-90,scale=0.5]{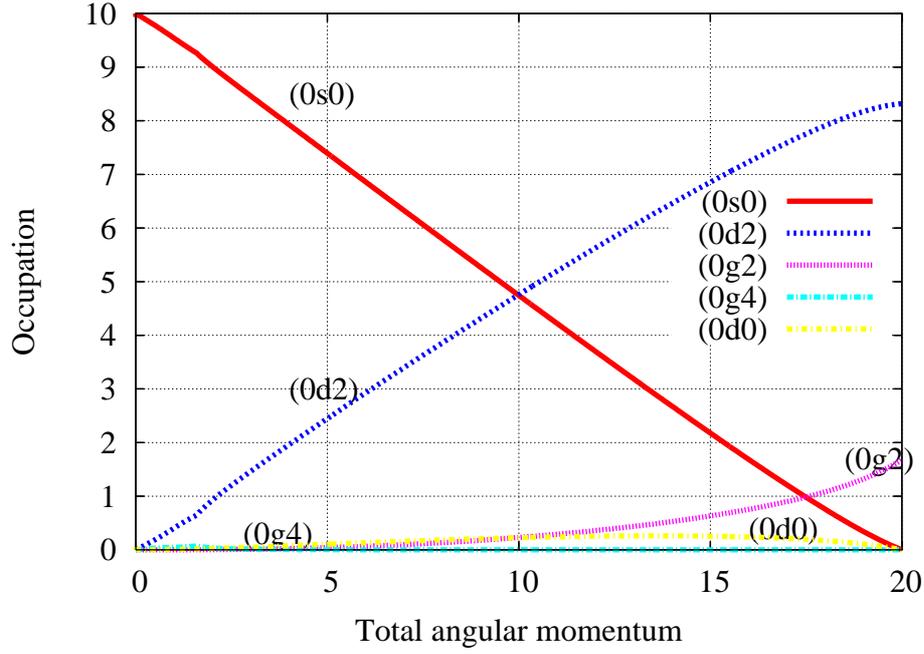}
\caption{(Color online) The occupation number $\rho_{\alpha\alpha}$ of the
 wave-function for  $\kappa/A^2=0.1$ (meV/$\mu$m$^4$).
The numbers in the figure ($n_\alpha l_\alpha m_\alpha$)
 stand for the principal, orbital and  magnetic quantum numbers of 
the harmonic oscillator state.}
\label{fig:occupation}
\end{figure*}

Figure \ref{fig:occupation} shows the occupation probability,
$\rho_{\alpha \alpha}$ for $\kappa/A^2=0.1$ meV/$\mu$m$^4$, where
$\rho_{\alpha \beta}$ is the density matrix defined as
\begin{equation}
\rho_{\alpha \beta} = \sum_{i} V_{\alpha i}^* V_{\beta i}.
\end{equation}
($V$ is a matrix appearing in the Bogoliubov-Valatin transformation,
Eq.(\ref{BV-trans}).) 
Although the occupation probabilities for 
$\kappa/A^ = 0.5, 1$ meV/$\mu$m$^4$ are not plotted in Figure 
\ref{fig:occupation},
we have calculated these occupation probabilities and found that
they are almost same as that of $\kappa/A^2 = 0.1$ meV/$\mu$m$^4$.
This result indicates that the wave-function does not strongly  
depend on the strength of the quadrupole-quadrupole interaction.
%

At low $L$, the $(0s0)$ states are the major component in the HFB state.
The higher the total angular momentum, the more the $(0d2)$ state 
admixes with the $(0s0)$ state. 
The $(0g2)$ component is also mixed at $L\simeq 2N=20$,
while the $(0s0)$ component vanishes.
This result suggests that
 the yrast state changes its structure gradually.

\begin{figure*}
\includegraphics[scale=0.8]{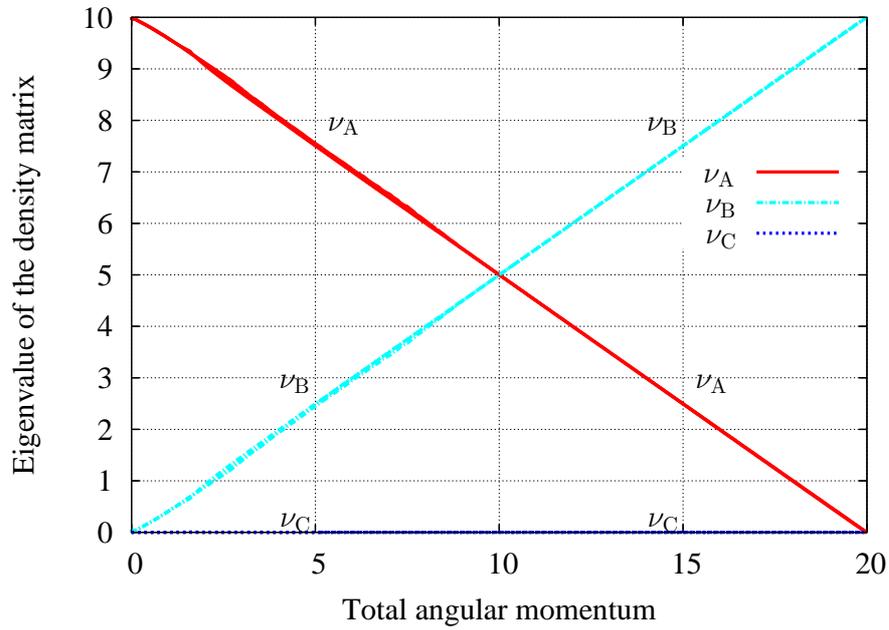}
\caption{(Color online) The eigenvalue $\nu_a$ of the density matrix 
$\rho_{\alpha \beta}$ for $\kappa/A^2 =0.1$
  (meV/$\mu$m$^4$) as a function of the total angular momentum. 
%
Only the three eigenvalues $\nu_{\rm A}, \nu_{\rm B}, \nu_{\rm C}$
are ploted because the other eigenvalues are almost zero. 
%
}
\label{fig:eigen}
\end{figure*}

Figure \ref{fig:eigen} shows the eigenvalue 
$\nu_a$ of the density matrix $\rho_{\alpha \beta}$ ,
as a function of $L$.
Only the plot for $\kappa/A^2 = 0.1$ (meV/$\mu$m$^4$) is displayed
because the occupation is almost independent of $\kappa/A^2$.
The largest eigenvalue in the figure 
is equal to the total particle number $N=10$, and
this situation happens only at $L=0$ and $2N (=20)$
In these cases, all the particles occupy only one single-particle state,
which can be regarded as the condensate state.
On the other hand, 
between $L=0$ and $2N$ ($0<L<2N=20$), 
the particles are shared by the two states $\psi^{\text{A}}$ and 
$\psi^{\text{B}}$, where $\nu_{\rm A} + \nu_{\rm B} = N = 10$.
This result indicates that most of the yrast states are non-condensates,
but a mixture of two single-particle components, $\psi^{\text{A}}$ and 
$\psi^{\text{B}}$.

\begin{figure*}
  \begin{tabular}{cc}
    \includegraphics[angle=-90,width=0.45\textwidth]{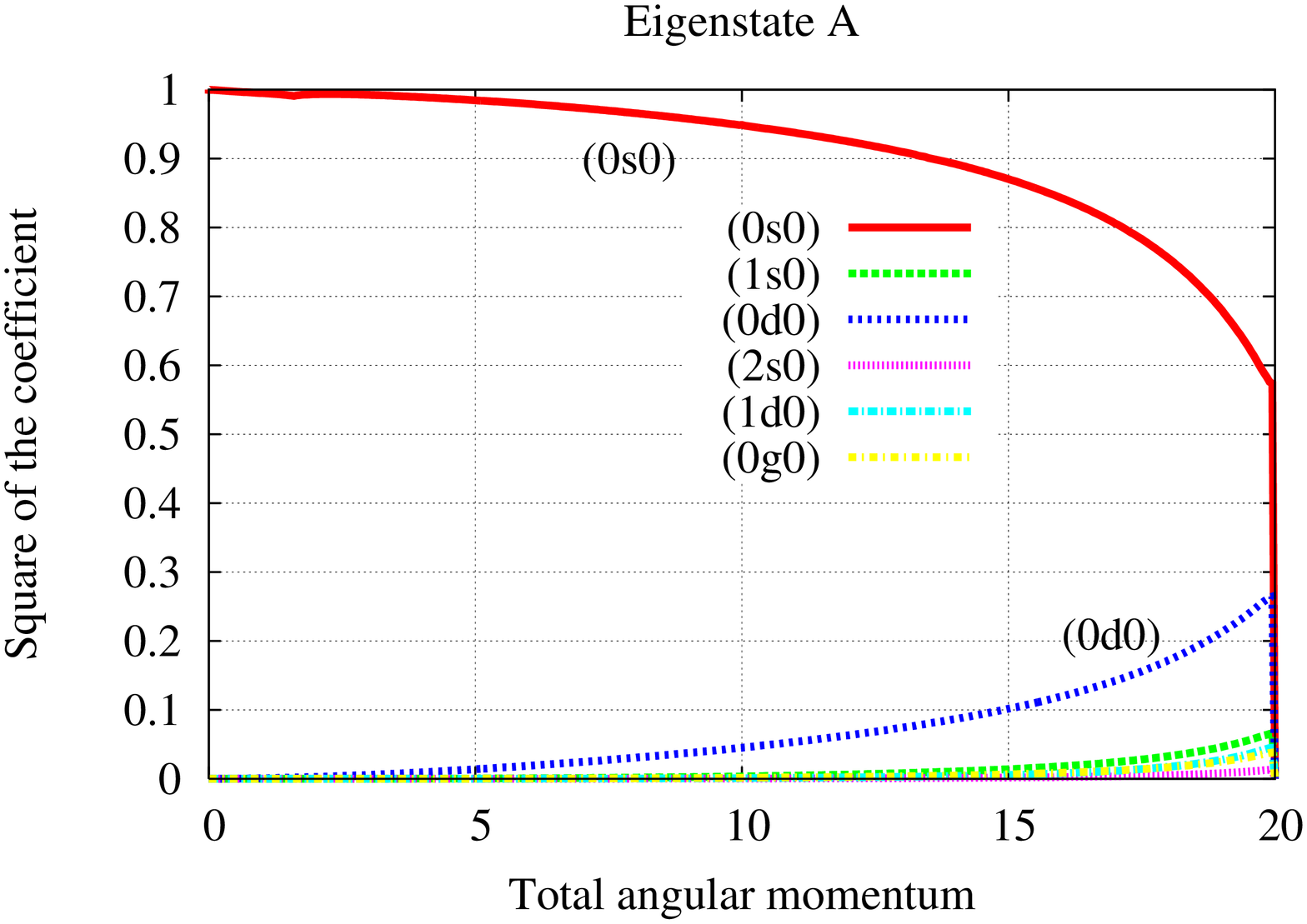} &
    \includegraphics[angle=-90,width=0.45\textwidth]{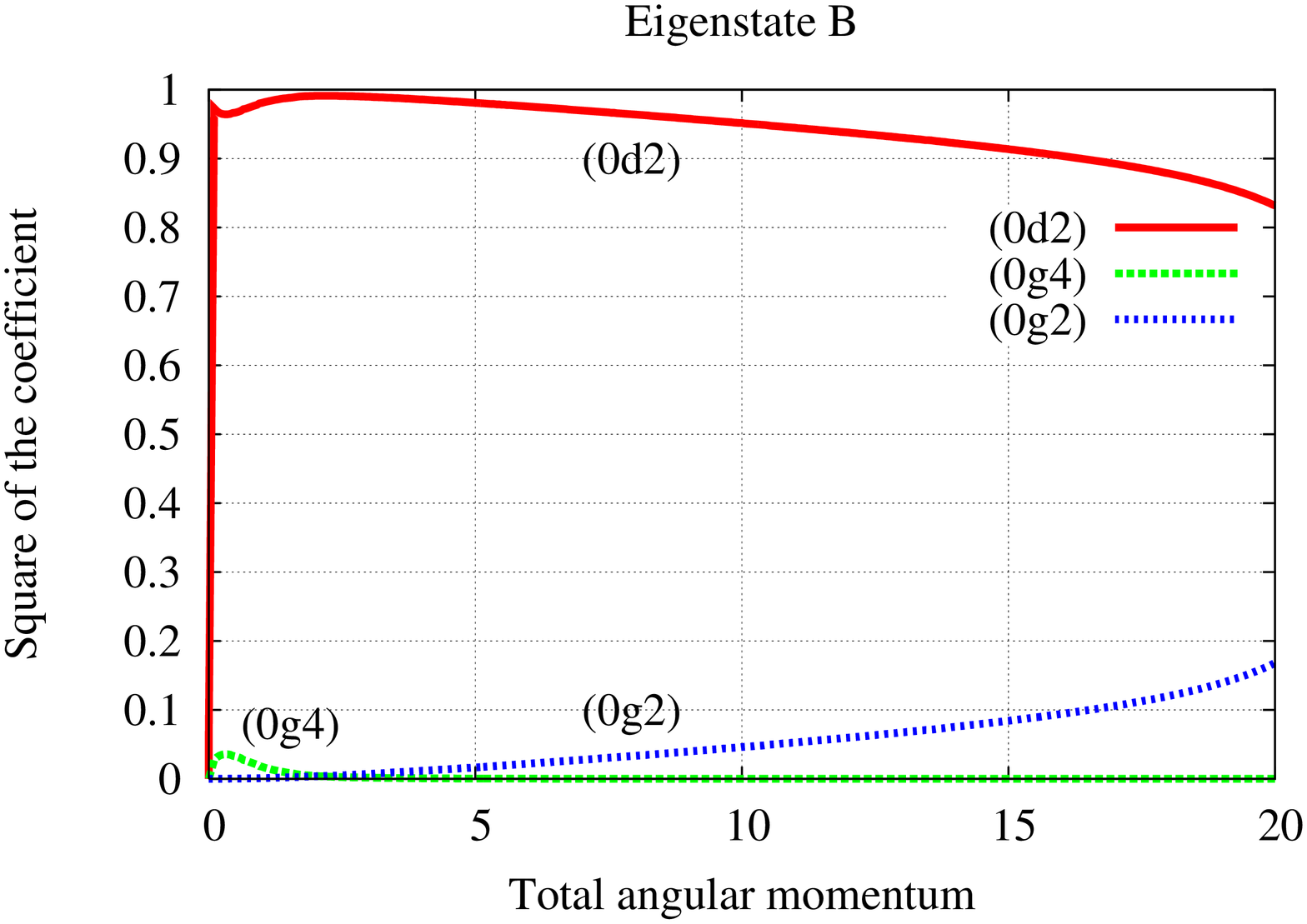}
  \end{tabular}
  \caption{(Color online) Left (right) panel represents the
    components for eigenstate $\psi^{\rm A}$ ($\psi^{\rm B}$) of the 
    density matrix as a function of the total angular momentum,
    respectively.
    The eigenstate $\psi^{\rm A}$ ($\psi^{\rm B}$)
    corresponds to the eigenvalue $\nu_{\rm A}$ ($\nu_{\rm B}$)
    in Figure \ref{fig:eigen}.
  }
\label{fig:component}
\end{figure*}

In Figure \ref{fig:component},
the eigenstates $\psi^{\text{A}}$ and $\psi^{\text{B}}$ are decomposed 
into the single-particle basis, and their components
are displayed in terms of probability $(v^a_\alpha)^2$, 
where $|\psi^a\rangle = \sum_\alpha v^a_\alpha |\alpha \rangle$.
First, as shown in the right panel of the figure, 
the state $\psi^{\text{A}}$ at $L=0$ 
is found to have a condensate structure into the $(0s0)$ state,
which is consistent with the occupation number calculation 
shown in Figure \ref{fig:occupation}.
As $L$ is increased, 
the $(0d0)$ component starts to mix with the $s$ component
although the contribution from the $d$ state is minor.
This mixture is consistent with the growth of the quadrupole deformation,
as shown in the left panel of Figure \ref{fig:betagamma}.
When $L\alt 5$, axial symmetry is broken around the cranking axis,
as shown in Figure \ref{fig:betagamma}. In this situation, the whole system
rotates in a collective manner, which is consistent with Mottelson's model
claiming that the yrast structure is dominated by the collective quadrupole
excitation at low $L$. Such a collective rotation was actually 
observed experimentally at lower rotational frequency 
before vortices are formed \cite{MIT}.
However, the linear dependence of $E$ on $L$ at low $L$ seen in 
Figure \ref{fig:energy} implies that the collective mode is
not the major mode in our model, but that the single-particle excitations are.
Next, in the right panel  of Figure \ref{fig:component}, 
the state $\psi^{\text{B}}$ is decomposed into the d and g states.
This is because the quadrupole-quadrupole interaction does not mix the states 
having different parity, so that the particles in the $s$ state 
can not be excited to the $p$ state, but the $d$ or $g$ states.
The major component is the $(0d2)$ state when angular momentum is small 
($L\alt 10$), whereas  the $(0g2)$ state starts to mix 
in the high angular momentum region ($L > 10$), 
due to the onset of deformation in the mean field.
Although the $(0g4)$ state is included already at $L\simeq 0$, 
this component can be considered as a minor component 
because the occupation $\nu_B$ is nearly 0 in this region. 

According to the previous discussion, at $L=2N$, 
the many-body state goes into the condensate in the state $\psi^{\text{B}}$, 
which is a mixture of the $(0d2)$ and $(0g2)$ states.
Both of these $(0d2)$ and $(0g2)$ states have angular dependence 
$\sin^2 \theta$, so that the density along the $z$-axis vanishes 
for these states. Therefore, the condensate at $L=2N$ is considered as 
a quantized vortex state carrying 2 ($\hbar$).

The HFB solutions obtained for the intermediate $L$ values ($0<L<20$) 
are quite different from the solutions obtained with  the GP equation. 
This is because the GP equation assumes 
an {\it a priori} existence of the condensate for the whole range of $L$,
which can be expressed as a Hartree state, $(a^{\dag}_0)^N|0\rangle$.
Although our HFB ansatz includes this condensate state as a special case,
the mathematical form for the HFB state is generally more complicated
to allow a linear combination form of multiple Hartree states,
in accordance with Eq. (\ref{HFBansatz}).
If the condensate is realized at any $L$,
all the particles should occupy the one single-particle state that
is generally expressed by a liner combination of the basis states, 
such as the $s$, $d$ and $g$ states.

As shown in Figure \ref{fig:eigen},
the yrast structure changes smoothly from the $\psi^{A}$-dominant states
to the $\psi^{B}$-dominant state, as $L$ is increased from 0 to $2N$.
Considering that the latter state at $L=2N$ corresponds to a
vortex state, a formation of the vortex starts as a shallow dent in the center
at small $L$.
As the amplitude of $\psi^{\text{B}}$ becomes larger for increasing $L$, 
the depth of the dent becomes deeper. 
Finally at $L=2N$, a complete vortex is formed, where density becomes zero. 
This mechanism of the vortex formation is very different from the one 
derived from the GP equation.
In the calculations using the GP equation \cite{KMP00}, 
a vortex enters from the ``outside''
of the system due to a continuity of the many-body wave function.
This sort of process needs a odd-number multipolarities, such as a dipole 
($\lambda=1$) and an octupole ($\lambda=3$) correlations, which are
missing from our present model.

These higher multipole correlations may play an important role 
also in a formation of vortex lattices, which violates rotational symmetry 
of the system. In the present framework, vortices appear always 
in the center of the system due to the axial symmetry possessed by the system.
To allow multiple vortices to appear away from the center, 
we need, at least, a mixing between these states of 
$(0s0)$, $(0d2)$ and $(0g2)$ to break the symmetry. 
This situation is realized only when $\langle Q_{2\pm 2}\rangle$, 
which is contained in the single-particle Hamiltonian, Eq.(\ref{eq-scf}), 
is non-zero. 
In other words, the deformation parameter $\gamma$ should not be equal 
to $0^{\circ}$ or $180^{\circ}$ to allow triaxial deformation bringing
anisotropy to the system. However, in the present calculations,
$\gamma$ takes the value of $180^{\circ}$ in a wide range of angular momentum 
($5 <L <20$). Consequently, the vortex lattice is not produced in 
our model.

\section{Summary}

Extending the 3D-cranked Hartree-Fock-Bogoliubov method,
we have performed the numerical calculation for a rotating many-boson system
interacting through a weak and repulsive interaction,
trapped inside an isotropic harmonic oscillator potential.
Unlike the Gross-Pitaevskii equation, our calculation does not assume 
the existence of the condensate {\it a priori}, but 
general many-body states in the framework of the HFB method.

We applied the method to a simple model where the two-body interaction
is chosen to be a separable type called the quadrupole-quadrupole interaction.
Parity is conserved for this interaction, so that only 
$\Delta l =2$ excitations is allowed through
the quadrupole-quadrupole interaction.

First of all, at $L=0$, our calculation shows that
the HFB state is interpreted as a condensate into $(0s0)$.

As increasing the total angular momentum $L$ from 0 to 20, 
the particles in the yrast state transfer from $\phi^A$ to $\phi^B$.
which have the magnetic quantum numbers $M_z=0$ and $M_z=2$, respectively. 
At low $L$, triaxial deformation is formed to allow collective rotation
around the cranking axis. However, the linearity $E\propto L$ implies 
that the major excitation mode is still single-particle excitations.
 At higher $L$, the system becomes oblate,
that is, axial symmetric around the rotating axis. Angular momentum is
thus produced by migrating from $s$ state to $d$ state.
These single-particle excitations are induced from the $s$ to
the $d$ state through the quadrupole-quadrupole interaction,
in agreement with the prediction by Mottelson.

Finally, at $L=2N$, the HFB state becomes another condensate,
in which all the particles occupy the single state expressed 
by a linear combination of the $(0d2)$ and $(0g2)$ states. 
This state can be interpreted 
as a vortex state having angular momentum $2 (\hbar)$.

In this way, the yrast structure changes gradually and smoothly
in our framework. 
This result can be accounted by the following two effects:
One is our choice of the HFB ansatz which allows a linear combination 
form of multiple Hartree states. The other is the finite-number effect of 
the total particle number ($N=10$).
These effects surely needs further investigations in the future studies.

We are currently proceeding to extend
our programming code to deal with more realistic inter-atomic potentials, and
plan to examine the effect of the pairing interaction in the HFB framework.

\section{Acknowledgment}
This work is supported by EPSRC with a grant EP/C520521/1.

\newpage

\end{document}